\documentclass{Interspeech2024}

\usepackage{multirow}
\usepackage{graphicx}
\usepackage{kotex}
\usepackage{multicol}
\usepackage{array}
\usepackage{subcaption}
\usepackage{cite}
\usepackage{algorithm}
\usepackage{algpseudocode}

\interspeechcameraready 

\title{High Fidelity Text-to-Speech Via Discrete Tokens Using Token Transducer and Group Masked Language Model}
\name[affiliation={1*}]{Joun Yeop}{Lee}
\name[affiliation={2*}]{Myeonghun}{Jeong}
\name[affiliation={2*}]{Minchan}{Kim}
\name[affiliation={1}]{Ji-Hyun}{Lee}
\name[affiliation={1}]{Hoon-Young}{Cho}
\name[affiliation={2}]{Nam Soo}{Kim}

\address{
  $^1$Samsung Research, Seoul, Republic of Korea\\
  $^2$Department of ECE and INMC,
  Seoul National University, Seoul, Republic of Korea}
\email{jounyeop.lee@samsung.com, \{mhjeong, mckim\}@hi.snu.ac.kr}

\keywords{speech synthesis, neural transducer, masked language modeling, semantic token, acoustic token}

\begin{document}

\maketitle
\def\thefootnote{*}\footnotetext{These authors contributed equally to this work.}

\begin{abstract}
    We propose a novel two-stage text-to-speech~(TTS) framework with two types of discrete tokens, i.e., semantic and acoustic tokens, for high-fidelity speech synthesis. It features two core components: the \textit{Interpreting} module, which processes text and a speech prompt into semantic tokens focusing on linguistic contents and alignment, and the \textit{Speaking} module, which captures the timbre of the target voice to generate acoustic tokens from semantic tokens, enriching speech reconstruction. The \textit{Interpreting} stage employs a transducer for its robustness in aligning text to speech. In contrast, the \textit{Speaking} stage utilizes a Conformer-based architecture integrated with a Grouped Masked Language Model~(G-MLM) to boost computational efficiency. Our experiments verify that this innovative structure surpasses the conventional models in the zero-shot scenario in terms of speech quality and speaker similarity.
    
\end{abstract}

\section{Introduction}
Recently, there has been a notable shift in Text-to-Speech (TTS) research towards the adoption of discrete speech tokens as intermediate features~\cite{kharitonov2023speak, wang2023neural, borsos2023soundstorm, du2023unicats}, which presents diverse options for model architecture and inference strategies. These tokens are broadly categorized into two types: semantic tokens and acoustic tokens, depending on their embedded information. Semantic tokens are typically derived through quantization applied to speech features containing contextualized linguistic details. These quantized speech features are sourced from various speech encoders such as self-supervised speech models~\cite{hsu2021hubert, baevski2020wav2vec} or speech recognition models~\cite{radford2023robust}. By encoding disentangled linguistic information into discrete codes, semantic tokens alleviate the complexities arising from acoustic diversity, thereby enabling a sharper focus on semantic content critical for enhancing intelligibility. In contrast, acoustic tokens represent codewords generated by neural codecs~\cite{zeghidour2021soundstream, defossez2023high ,yang2023hifi, kumar2024high}, which have witnessed significant advancements in recent years. These tokens encapsulate acoustic details of raw waveforms, serving as alternatives to traditional frame-level acoustic features such as mel-spectrograms.

Incorporating discrete tokens into TTS offers numerous advantages over conventional speech modeling in the continuous domain. Above all, targeting discrete tokens simplifies the representation of one-to-many mappings by facilitating a categorical distribution for the output space, addressing the complex challenges posed by continuous domain generative modeling. Moreover, the discrete output space enables the integration of various specialized schemes. Particularly noteworthy is its ability to leverage recent advancements in large language models~(LLM). For example, SPEAR-TTS\cite{kharitonov2023speak} and VALL-E~\cite{wang2023neural} employ in-context learning for prompt-oriented zero-shot TTS, complemented by speech continuation tasks. Additionally, recent progress in iterative sampling with masked language models (MLM)~\cite{borsos2023soundstorm, jeong2024efficient, chang2022maskgit} has demonstrated high-fidelity speech synthesis with parallel computation, a capability confined to discrete sequences. Furthermore, the discrete output space can facilitate the development of robust alignment modeling, simplifying the adoption of transducers\cite{kim2024utilizing, du2024vall} within the TTS framework.

In this work, we propose a high-fidelity TTS framework designed to optimize the use of semantic and acoustic tokens, leveraging the benefits of discrete tokenization. This framework follows a two-stage procedure: converting text into semantic tokens~(\textit{Interpreting}) and then into acoustic tokens~(\textit{Speaking}). There have been attempts to use such a two-stage architecture with discrete tokens. For example, SPEAR-TTS uses ``reading" and ``speaking" modules similar to ours. However, the ``reading" utilizes an encoder-decoder transformer structure that cannot guarantee a monotonic alignment path, and has some complicated training methods that are hard to reproduce. Also, ``speaking" uses a huge decoder-only transformer structure which results in computational inefficiency induced by sequential generation. On the contrary, in our architecture, the \textit{Interpreting} stage employs a transducer~\cite{graves2012sequence} for robust alignment modeling, capitalizing on the inherent monotonic alignment constraint in seq2seq models for efficient computation and alignment modeling. Also, in the \textit{Speaking} stage, we employ an MLM with parallel iterative sampling for acoustic token generation, leveraging the pre-established alignment between semantic and acoustic tokens to enhance inference speed and efficiency. Specifically, we utilize a group masked language model~\cite{jeong2024efficient} with HiFi-Codec~\cite{yang2023hifi} to reduce sampling iterations and improve speech quality. Both phases efficiently exploit conditional dependencies between output sequences during inference, a crucial factor for our high-fidelity speech generation. Experimental results on zero-shot TTS demonstrate the superiority of our model over baseline models in terms of speech quality and speaker similarity. Audio samples are available on our demo page\footnote{https://srtts.github.io/interpreting-speaking/}.

\section{Method}

\begin{figure*}
 \centering
 \includegraphics[width=1.9\columnwidth]{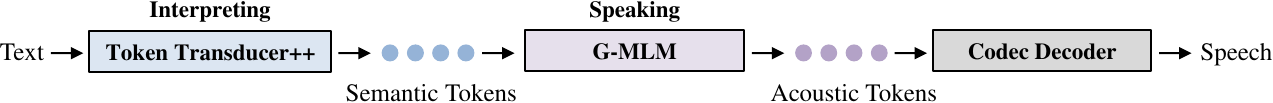}%
 \caption{Overall architecture of the proposed model.} 
 \label{figure1}
\end{figure*}

The proposed TTS framework consists of a text-to-semantic token stage, named \textit{Interpreting}, and semantic to the acoustic token stage, named \textit{Speaking}. Then, the acoustic tokens are converted into raw wave through a codec decoder. The entire framework is illustrated in Figure~1.

As semantic tokens, we exploit the index sequence of k-means clustering on wav2vec~2.0~\cite{baevski2020wav2vec} embeddings, akin to \cite{kim2024utilizing}. According to \cite{kim2024utilizing}, semantic tokens mainly focus on phonetic information, but it also dealing with some prosodic information such as speech rate and overall pitch contour. For acoustic token extraction, we leverage the HiFi-Codec~\cite{yang2023hifi}, a sophisticated neural codec tailored for TTS applications. The HiFi-Codec employs the Group-Residual Vector Quantization~(G-RVQ) to derive multiple streams of discrete token sequences. Specifically, we use the bi-group and bi-depth G-RVQ to maintain the balance between performance and computational efficiency. We denote the semantic token at time step~$t$, as $s_t$, while the acoustic token is represented as $y_t$, each comprising $y_t^{i,j}$, with $i,j\in\{0,1\}$ denoting indices corresponding to the group and depth of G-RVQ, respectively.

\subsection{Interpreting}
In the \textit{Interpreting} stage, we generate a semantic token sequence~$\textbf{s}_{0:T}$ from the input text~$\textbf{x}_{1:N}$, addressing alignment between text and semantic tokens, as well as controlling prosody embedded in semantic tokens. While attention-based language models are often considered for this seq2seq translation task, they cannot exploit the inherent monotonic constraint of alignment, making them susceptible to misalignment issues and requiring computationally intensive key-query matching processes. Instead, we adopt a transducer, referred to as Token Transducer++, a modified version of the Token Transducer of \cite{kim2024utilizing}. Transducers are specifically designed architecture for discrete seq2seq models with monotonic alignment constraint, achieved through alignment lattice~(Figure 2.~(a)) and a special blank token~$\varnothing$, which indicates the transition to the next input frame~(horizontal arrows in Figure 2.~(a)). 
The Token Transducer++ follows the same transducer formulation of \cite{kim2024utilizing}, where the training objective is expressed as:  
\begin{equation} \label{eq:loss1}
\begin{split}
  \mathcal{L}_{inter} &= -\log{P(\textbf{s}|\textbf{x})}\\
  &=-\log{\sum_{\mathcal{A} \in \mathcal{F}^{-1}(\textbf{s})}{P(\mathcal{A}|\textbf{x}, \textbf{p}_{s})}}.
\end{split}
\end{equation}
Here, $\mathcal{A}$ represents a possible monotonic path and, $\mathcal{F}^{-1}$ denotes the inverse of the blank removal function $\mathcal{F}$, indicating the marginalization over all $\mathcal{A}$. Also, $\textbf{p}_{s}$ denotes the speech prompt for the semantic token generation, i.e., semantic prompt, which controls paralinguistic information embedded in semantic token sequence such as speech rate and pitch contour.

As depicted in Figure~2 (a), the Token Transducer++ has a modularized structure comprising four components: a text encoder, a reference encoder, a prediction network, and a joint network. The text encoder processes text input, producing a sequence of text embeddings. The reference encoder encodes speech prompt into fixed-dimensional reference embedding. The prediction network operates in an autoregressive manner, leveraging preceding semantic tokens and the reference embedding to generate the semantic token embedding for the current frame. Lastly, the joint network integrates text and semantic token embeddings, generating output probabilities, crucial for lattice construction. Notably, the Token Transducer++ mirrors the architecture of each module detailed in \cite{kim2024utilizing}. However, there are two drawbacks of transducers: (1) the autoregressive nature of the joint network presents a significant computational bottleneck during inference, and (2) the joint network's frame-wise computation neglects temporal context. To overcome these problems, when compared to the original Token Transducer, we largely reduce the size of the joint network and inject reference embedding to the prediction network instead of the joint network for temporal consideration of reference embedding. We add the reference embedding to the input embedding of the prediction network. These simple modifications not only boost inference speed but also enhance overall performance.

\begin{figure}
 \centering
 \begin{subfigure}{0.7\columnwidth}
 \includegraphics[width=\columnwidth]{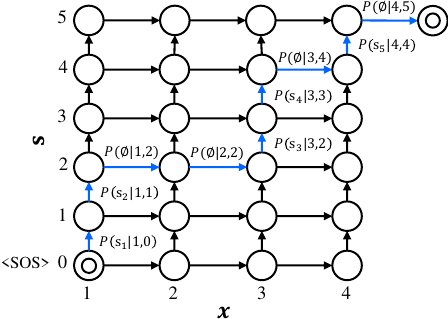}%
 \caption{Alignment lattice of a transducer}%
 \label{subfiga}%
 \end{subfigure}\vspace{0.3cm}
 \begin{subfigure}{0.9\columnwidth}
 \includegraphics[width=\columnwidth]{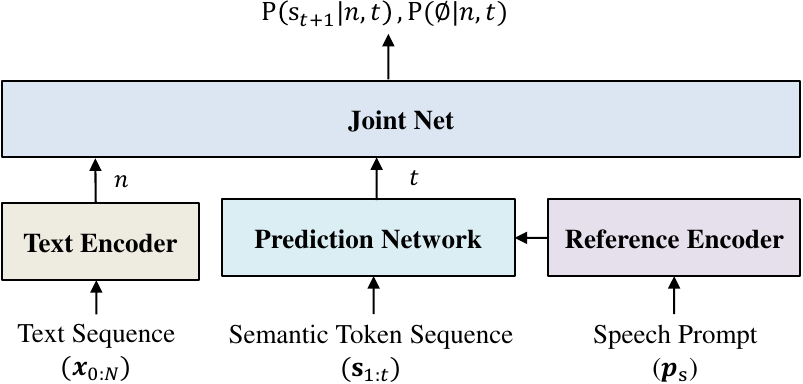}%
 \caption{Architecture of Token Transducer++}
 \label{subfigb}%
 \end{subfigure}
 \caption{Interpreting procedure} 
 \label{figure1}
\end{figure}

\subsection{Speaking}
The \textit{Speaking} stage aims to translate semantic tokens produced in the \textit{Interpreting} stage into acoustic tokens, utilizing a prompt for acoustic guidance. We tackle this pre-aligned seq2seq task via Masked Language Modeling~(MLM), which effectively incorporates prompt speaker information through in-context learning. To achieve this, the \textit{Speaking} module is meticulously trained to optimize the prediction of masked acoustic tokens as follows:

\begin{equation} \label{eq:loss2}
  \mathcal{L}_{speak}=-\sum_{\forall y \in \textbf{Y}_M}{P(y|\textbf{Y}_U, \textbf{s}, \textbf{p}_{a})},
\end{equation}
where $\textbf{Y}_M$, $\textbf{Y}_U$ represents masked and unmasked target acoustic tokens. Also, $\textbf{p}_a$ denotes an acoustic prompt, and it provides detailed acoustic attributes~(i.e., timbre, acoustic condition) of the prompt speaker.

Given the characteristics of RVQ-based acoustic tokens, there have been MLM approaches that capture both of temporal and RVQ-level-wise conditional dependency~\cite{borsos2023soundstorm, garcia2023vampnet}. Among these methods, we employ the Group-MLM~(G-MLM) approach in our \textit{Speaking} module, which is specifically designed for G-RVQ acoustic tokens. By simultaneously masking tokens at the same level across different groups, it additionally captures group-wise conditional dependency. Such a masking strategy enables the model to more easily predict masked tokens, ultimately facilitating efficient decoding. The inference method that mimics this approach is illustrated in Figure~3.~(a). First, coarse-grained acoustic tokens from different groups (e.g., $\mathbf{y}^{0,0}$, $\mathbf{y}^{1,0}$) are obtained through $N_c$ iterations of iterative parallel decoding. Then, fine-grained acoustic tokens are predicted all at once. Note that this sampling scheme, named Group-Iterative Parallel Decoding (G-IPD), which originates from the masking strategy of G-MLM, improves the audio quality even with a small number of iterations.

\begin{figure}
 \centering
 \begin{subfigure}{0.9\columnwidth}
 \includegraphics[width=\columnwidth]{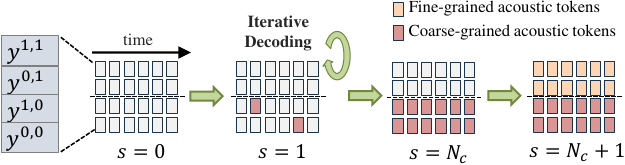}%
 \caption{Group-Iterative Parallel Decoding~(G-IPD)}%
 \label{subfiga}%
 \end{subfigure}\vspace{0.3cm}
 \begin{subfigure}{0.9\columnwidth}
 \includegraphics[width=\columnwidth]{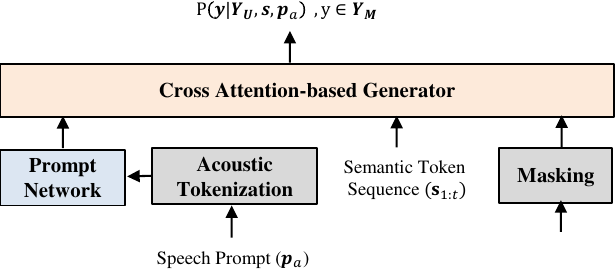}%
 \caption{Architecture of Speaking}%
 \label{subfigb}%
 \end{subfigure}
 \caption{Speaking procedure} 
 \label{figure1}
\end{figure}

As illustrated in Figure~3.~(b), the \textit{Speaking} part comprises a prompt network and a generator based on cross-attention mechanisms. Each module follows the architecture described in~\cite{jeong2024efficient}. The prompt network and generator are based on a bidirectional conformer structure~\cite{gulati20_interspeech} to learn the underlying contextual information. The prompt embedding processed from the prompt network operates as the key and value, while the aggregated embeddings of the semantic tokens and partially masked acoustic tokens serve as the query. This cross-attention-based architecture has been proven to be efficient in terms of computational cost and inference speed, as it allows for caching of key and value during iterative decoding. This structure streamlines the whole inference process while offering rich prompt information~\cite{shen2024naturalspeech, jeong2024efficient}.

\subsection{Discussion on Framework Design}

The division of the two stages offers several distinct advantages. Firstly, each stage can concentrate on different aspects of TTS objective. The \textit{Interpreting} stage can solely prioritize alignment modeling and linguistic accuracy, whereas the \textit{Speaking} stage can focus more on high-fidelity and handling one-to-many mappings due to acoustic diversity. We design sophisticated architectures tailored to each role, considering both speech quality and inference efficiency. Additionally, from a data perspective, the \textit{Interpreting} stage relies on text and speech pair data, whereas the \textit{Speaking} stage doesn't require text annotations. This flexibility enables us to leverage unlabeled data, which is more abundant, resulting in higher speech quality and the ability to represent diverse acoustic conditions. Moreover, regarding zero-shot adaptation, we can separately control each stage as they govern different aspects of paralinguistic information. The \textit{Interpreting} stage manages speech rate and global prosodic dynamics, while the \textit{Speaking} stage focuses on timbre and acoustic attributes. This separation provides us with the flexibility to independently control each component using different speech prompts, thereby enhancing overall controllability across a broader spectrum.

\section{Experiments}
\subsection{Experimental Setting}
\textbf{Dataset:}
We conducted zero-shot TTS experiments using the LibriTTS~\cite{zen2019libritts} corpus. We used all of the training subsets~(train-clean-100, train-clean-360, train-other-500) for training and the test-clean subset for evaluation. All the sentences were phonemized to International Phonetic Alphabet (IPA) using the phonemizer library\footnote{\url{https://github.com/bootphon/phonemizer}}.\\
\textbf{Implementation Details:}
Following~\cite{kim2024utilizing}, we employed semantic tokens obtained through $k$-means clustering on the wav2vec 2.0-XLSR model~\cite{baevski2020wav2vec}\footnote{\url{https://huggingface.co/facebook/wav2vec2-xlsr-53-espeak-cv-ft}}, with $k$ set to 512. The acoustic tokenization was conducted using the pre-trained HiFi-Codec\footnote{\url{https://github.com/yangdongchao/AcademiCodec}}, optimized for 24kHz speech samples and a 320 times down-sampling rate. We constructed the Token Transducer++ with same the configuration as the Token Transducer in \cite{kim2024utilizing}, but we reduced the number of feed-forward blocks from 3 to 1. The Token Transducer++ were trained for 30 epochs with dynamic batch size containing up to 240 seconds on two Quadro RTX 8000 GPUs. The implementation of \textit{Speaking} stage followed \cite{jeong2024efficient}, unless specified otherwise. The G-MLM underwent training for $350k$ iterations with a batch size of 128 utilizing four NVIDIA A100 GPUs. We set $N_c$ to 16, resulting in a total of 17 iterations for sampling.\\
\textbf{Baselines:} We built three baseline models for comparison: VITS, VALLE-X~\cite{zhang2023speak}, and Kim et al.~\cite{kim2024utilizing}. To adapt the baseline VITS to the zero-shot adaptive scenario, we incorporated the ECAPA-TDNN~\cite{desplanques20_interspeech} structure as a reference encoder. We used the open-source implementation for VALLE-X~\footnote{\url{https://github.com/Plachtaa/VALL-E-X}}\\
\textbf{Evaluation Metrics:}
For objective evaluations, we assessed the character error rate (CER) using a pretrained Whisper large model~\cite{radford2023robust}, leveraging the official implementation\footnote{\url{https://github.com/openai/whisper}}. Additionally, we conducted averaged speaker embedding cosine similarity~(SECS) analysis to evaluate speaker similarity between speech prompts and synthesized samples, utilizing the pre-trained WavLM large speaker verification model\footnote{\url{https://github.com/microsoft/UniSpeech/tree/main/downstreams/speaker_verification}}. We randomly selected 500 utterances from the test dataset for these objective assessments.

For subjective evaluation, we measured Mean Opinion Score (MOS) and Similarity MOS (SMOS). The MOS test evaluated perceptual speech quality as rated by testers. In the SMOS test, testers rated the similarity of speaker and prosody between speech prompts and synthesized samples. Both tests utilized a scoring range from 1 to 5, with higher scores indicating better performance. For both subjective tests, 90 testers participated via Amazon MTurk\footnote{ \url{https://www.mturk.com}} platform.

\subsection{Results: Zero-Shot Multi-Speaker TTS}
According to Table 1, our proposed model exhibited superior performance compared to the baselines across all assessments. Notably, VALLE-X demonstrated the worst CER due to misalignment issues stemming from the lack of a monotonic alignment constraint. In contrast, alignment models based on the transducer approach, including \cite{kim2024utilizing} and the proposed model, exhibited higher intelligibility owing to their robust alignment capabilities. Furthermore, when compared to the transducer-based model~\cite{kim2024utilizing}, our proposed model presented enhanced speech quality and speaker similarity, attributed to the G-MLM utilized in the \textit{Speaking} stage. Despite some inconsistencies in the results, it is noteworthy that our proposed architecture outperformed all assessed perspectives.

\begin{table}
\caption{Results of zero-shot TTS. MOS and SMOS are represented with 95\% confidence intervals.}
\label{}
\centering
\begin{tabular}{l c c c c}
\toprule
\textbf{Method} & \textbf{MOS}& \textbf{SMOS} & \textbf{CER}& \textbf{SECS} \\ 
\midrule
Ground Truth & 4.21\footnotesize{$\pm$0.09} &4.13\footnotesize{$\pm$0.10} & 0.97 & 0.653  \\
\midrule
VITS      &3.52\footnotesize{$\pm$0.16} & 3.23\footnotesize{$\pm$0.20} & 4.81 & 0.385    \\
VALLE-X     &3.75\footnotesize{$\pm$0.14} & 3.08\footnotesize{$\pm$0.21}  & 8.72 & 0.425  \\
Kim et al.~\cite{kim2024utilizing}     &3.69\footnotesize{$\pm$0.16} & 3.55\footnotesize{$\pm$0.21}  & 2.47 & 0.467 \\
\midrule
Proposed   &\textbf{3.94\footnotesize{$\pm$0.14}} & \textbf{3.64\footnotesize{$\pm$0.19}}   & \textbf{2.34} & \textbf{0.512}     \\
\bottomrule
\end{tabular}
\end{table}
\subsection{Ablation: TokenTransducer++}
To assess the improvement brought by Token Transducer++ over the original Token Transducer~\cite{kim2024utilizing}, we conducted a comprehensive comparative analysis employing the same proposed $Speaking$ stage. We compared objective measures including real-time factor~(RTF) of the Token Transducer inference, CER and SECS. Additionally, we computed the negative log-likelihood (NLL), employing the same methodology as the training objective, for both transducers. As detailed in Table 2, Token Transducer++ demonstrates significantly improved inference speed compared to the Token Transducer, even with better CER and NLL scores. While Token Transducer++ exhibits marginally lower SECS, we note that SECS is notably influenced by the \textit{Speaking} stage, as evidenced by comparisons with \cite{kim2024utilizing} presented in Table 1. These results underscore the efficacy of prosody conditioning within the prediction network, even with a more compact joint network architecture. This integration not only accelerated inference speed but also yields overall performance enhancements.

\begin{table}[h]
\caption{Comparison of performance between the Token Transducer and the Token Transducer++. The RTF is calculated on RTF8000 GPU.}
\label{}
\centering
\begin{tabular}{l c c c c}
\toprule
\textbf{Method} & \textbf{RTF} & \textbf{CER} & \textbf{SECS} & \textbf{NLL}\\ 
\midrule
Token Transducer      &14.35 & 2.50 & \textbf{0.518} & 1.326     \\
Token Transducer++      &\textbf{23.38} & \textbf{2.34} & 0.512 & \textbf{1.300}     \\
\bottomrule
\end{tabular}
\end{table}

\subsection{Ablation: Prosody Controllability}
We investigated the controllability using different speech prompts for both the \textit{Interpreting} and \textit{Speaking} stages. Following the approach outlined in \cite{kim2024utilizing}, we generated samples under two conditions: (1) when $\textbf{p}_{s}$ equals $\textbf{p}_{a}$, and (2) when $\textbf{p}_{s}$ differs from $\textbf{p}_{a}$ (selected from different speakers). The former represents the general zero-shot TTS scenario, while the latter assesses the ability to independently control semantic paralinguistic elements (such as speech rate and prosody) and acoustic conditions (including speaker identity, timbre, and environmental factors). This separation allows for disentangled prosody controllability. In Table 3, we computed the speaker embedding cosine similarity (SECS) between the generated speech and the speech prompts $\textbf{p}_{s}$ and $\textbf{p}_{a}$, comparing the proposed method with different \textit{Speaking} stage implementations, using the speech generator from \cite{kim2024utilizing}. Across all cases, our proposed model exhibited higher scores, indicating that the proposed G-MLM-based \textit{Speaking} approach offers superior speaker similarity in both scenarios. The results shows speaker similarity is mostly controlled by \textit{Speaking} stage~($\textbf{p}_{a}$), while linguistic prosody is controlled by semantic token generation~($\textbf{p}_{s}$)~\cite{kim2024utilizing}. For subjective verification, we uploaded samples illustrating the separated control scenario to our demo page.

\begin{table}[h]
\centering
\caption{Comparison of SECS between generated samples and the speech prompts. The $\textbf{p}_{rand}$ denotes arbitrary samples from the test set, which are used as the standard value.}
\label{tab:my-table}

\begin{tabular}{@{}cccc@{}}
\toprule
\multicolumn{2}{c}{\multirow{2}{*}{\textbf{Prompt}}} & \multicolumn{2}{c}{\textbf{SECS}} \\ \cmidrule(l){3-4} 
\multicolumn{2}{c}{}                           & Kim et al.~\cite{kim2024utilizing}      & Proposed      \\ \midrule
\multicolumn{2}{c}{$\textbf{p}_{rand}$}                         & 0.116        & \textbf{0.122}             \\ \midrule
\multicolumn{2}{c}{$\textbf{p}_{s}=\textbf{p}_{a}$}                         & 0.462        & \textbf{0.514}             \\ \midrule
\multirow{2}{*}{$\textbf{p}_{s}\neq\textbf{p}_{a}$} & $\textbf{p}_{s}$   & 0.119       & \textbf{0.128}             \\ \cmidrule(l){2-4}  &$\textbf{p}_{a}$            & 0.432        & \textbf{0.481}             \\
\bottomrule
\end{tabular}%
\end{table}


\section{Conclusions}
In this paper, we introduced a two-stage text-to-speech~(TTS) system designed to achieve high-fidelity speech synthesis through the utilization of semantic and acoustic tokens. The first stage, termed the \textit{Interpreting} module, effectively processes text and a speech prompt into semantic tokens, ensuring precise pronunciation and alignment. Following this, the \textit{Speaking} module takes over, employing these semantic tokens to generate acoustic tokens that capture the target voice's acoustic attribute~(timbre, acoustic condition), significantly enhancing the speech reconstruction process. Experimental results verify that our proposed method outperforms state-of-the-art baselines regarding speech intelligibility, audio quality, and speaker similarity. For future work, we plan to extend our proposed framework to multiple languages and make our model encompass a broader range of speech-generation tasks, including singing voice synthesis. Also, we will enlarge the training dataset to the unlabeled speech data to increase the generalization of \textit{Speaking}.

\bibliographystyle{IEEEtran}
\bibliography{mybib}

\begin{thebibliography}{10}
\providecommand{\url}[1]{#1}
\csname url@samestyle\endcsname
\providecommand{\newblock}{\relax}
\providecommand{\bibinfo}[2]{#2}
\providecommand{\BIBentrySTDinterwordspacing}{\spaceskip=0pt\relax}
\providecommand{\BIBentryALTinterwordstretchfactor}{4}
\providecommand{\BIBentryALTinterwordspacing}{\spaceskip=\fontdimen2\font plus
\BIBentryALTinterwordstretchfactor\fontdimen3\font minus \fontdimen4\font\relax}
\providecommand{\BIBforeignlanguage}[2]{{%
\expandafter\ifx\csname l@#1\endcsname\relax
\typeout{** WARNING: IEEEtran.bst: No hyphenation pattern has been}%
\typeout{** loaded for the language `#1'. Using the pattern for}%
\typeout{** the default language instead.}%
\else
\language=\csname l@#1\endcsname
\fi
#2}}
\providecommand{\BIBdecl}{\relax}
\BIBdecl

\bibitem{kharitonov2023speak}
E.~Kharitonov, D.~Vincent, Z.~Borsos, R.~Marinier, S.~Girgin, O.~Pietquin, M.~Sharifi, M.~Tagliasacchi, and N.~Zeghidour, ``Speak, read and prompt: High-fidelity text-to-speech with minimal supervision,'' \emph{Transactions of the Association for Computational Linguistics}, vol.~11, pp. 1703--1718, 2023.

\bibitem{wang2023neural}
C.~Wang, S.~Chen, Y.~Wu, Z.~Zhang, L.~Zhou, S.~Liu, Z.~Chen, Y.~Liu, H.~Wang, J.~Li \emph{et~al.}, ``Neural codec language models are zero-shot text to speech synthesizers,'' \emph{arXiv preprint arXiv:2301.02111}, 2023.

\bibitem{borsos2023soundstorm}
Z.~Borsos, M.~Sharifi, D.~Vincent, E.~Kharitonov, N.~Zeghidour, and M.~Tagliasacchi, ``Soundstorm: Efficient parallel audio generation,'' \emph{arXiv preprint arXiv:2305.09636}, 2023.

\bibitem{du2023unicats}
C.~Du, Y.~Guo, F.~Shen, Z.~Liu, Z.~Liang, X.~Chen, S.~Wang, H.~Zhang, and K.~Yu, ``Unicats: A unified context-aware text-to-speech framework with contextual vq-diffusion and vocoding,'' \emph{arXiv preprint arXiv:2306.07547}, 2023.

\bibitem{hsu2021hubert}
W.-N. Hsu, B.~Bolte, Y.-H.~H. Tsai, K.~Lakhotia, R.~Salakhutdinov, and A.~Mohamed, ``Hubert: Self-supervised speech representation learning by masked prediction of hidden units,'' \emph{IEEE/ACM Transactions on Audio, Speech, and Language Processing}, vol.~29, pp. 3451--3460, 2021.

\bibitem{baevski2020wav2vec}
A.~Baevski, Y.~Zhou, A.~Mohamed, and M.~Auli, ``wav2vec 2.0: A framework for self-supervised learning of speech representations,'' \emph{Advances in neural information processing systems}, vol.~33, pp. 12\,449--12\,460, 2020.

\bibitem{radford2023robust}
A.~Radford, J.~W. Kim, T.~Xu, G.~Brockman, C.~McLeavey, and I.~Sutskever, ``Robust speech recognition via large-scale weak supervision,'' in \emph{International Conference on Machine Learning}.\hskip 1em plus 0.5em minus 0.4em\relax PMLR, 2023, pp. 28\,492--28\,518.

\bibitem{zeghidour2021soundstream}
N.~Zeghidour, A.~Luebs, A.~Omran, J.~Skoglund, and M.~Tagliasacchi, ``Soundstream: An end-to-end neural audio codec,'' \emph{IEEE/ACM Transactions on Audio, Speech, and Language Processing}, vol.~30, pp. 495--507, 2021.

\bibitem{defossez2023high}
A.~D{\'e}fossez, J.~Copet, G.~Synnaeve, and Y.~Adi, ``High fidelity neural audio compression,'' \emph{Transactions on Machine Learning Research}, 2023, featured Certification, Reproducibility Certification.

\bibitem{yang2023hifi}
D.~Yang, S.~Liu, R.~Huang, J.~Tian, C.~Weng, and Y.~Zou, ``Hifi-codec: Group-residual vector quantization for high fidelity audio codec,'' \emph{arXiv preprint arXiv:2305.02765}, 2023.

\bibitem{kumar2024high}
R.~Kumar, P.~Seetharaman, A.~Luebs, I.~Kumar, and K.~Kumar, ``High-fidelity audio compression with improved rvqgan,'' \emph{Advances in Neural Information Processing Systems}, vol.~36, 2024.

\bibitem{jeong2024efficient}
M.~Jeong, M.~Kim, J.~Y. Lee, and N.~S. Kim, ``Efficient parallel audio generation using group masked language modeling,'' \emph{arXiv preprint arXiv:2401.01099}, 2024.

\bibitem{chang2022maskgit}
H.~Chang, H.~Zhang, L.~Jiang, C.~Liu, and W.~T. Freeman, ``Maskgit: Masked generative image transformer,'' in \emph{Proceedings of the IEEE/CVF Conference on Computer Vision and Pattern Recognition}, 2022, pp. 11\,315--11\,325.

\bibitem{kim2024utilizing}
M.~Kim, M.~Jeong, B.~J. Choi, S.~Kim, J.~Y. Lee, and N.~S. Kim, ``Utilizing neural transducers for two-stage text-to-speech via semantic token prediction,'' \emph{arXiv preprint arXiv:2401.01498}, 2024.

\bibitem{du2024vall}
C.~Du, Y.~Guo, H.~Wang, Y.~Yang, Z.~Niu, S.~Wang, H.~Zhang, X.~Chen, and K.~Yu, ``Vall-t: Decoder-only generative transducer for robust and decoding-controllable text-to-speech,'' \emph{arXiv preprint arXiv:2401.14321}, 2024.

\bibitem{graves2012sequence}
A.~Graves, ``Sequence transduction with recurrent neural networks,'' \emph{in Representation Learning Worksop, ICML}, 2012.

\bibitem{garcia2023vampnet}
H.~F.~F. Garcia, P.~Seetharaman, R.~Kumar, and B.~Pardo, ``Vampnet: Music generation via masked acoustic token modeling,'' in \emph{Ismir 2023 Hybrid Conference}, 2023.

\bibitem{gulati20_interspeech}
A.~Gulati, J.~Qin, C.-C. Chiu, N.~Parmar, Y.~Zhang, J.~Yu, W.~Han, S.~Wang, Z.~Zhang, Y.~Wu, and R.~Pang, ``{Conformer: Convolution-augmented Transformer for Speech Recognition},'' in \emph{Proc. Interspeech 2020}, 2020, pp. 5036--5040.

\bibitem{shen2024naturalspeech}
K.~Shen, Z.~Ju, X.~Tan, E.~Liu, Y.~Leng, L.~He, T.~Qin, sheng zhao, and J.~Bian, ``Naturalspeech 2: Latent diffusion models are natural and zero-shot speech and singing synthesizers,'' in \emph{The Twelfth International Conference on Learning Representations}, 2024.

\bibitem{zen2019libritts}
H.~Zen, V.~Dang, R.~Clark, Y.~Zhang, R.~J. Weiss, Y.~Jia, Z.~Chen, and Y.~Wu, ``Libritts: A corpus derived from librispeech for text-to-speech,'' \emph{Interspeech 2019}, 2019.

\bibitem{zhang2023speak}
Z.~Zhang, L.~Zhou, C.~Wang, S.~Chen, Y.~Wu, S.~Liu, Z.~Chen, Y.~Liu, H.~Wang, J.~Li \emph{et~al.}, ``Speak foreign languages with your own voice: Cross-lingual neural codec language modeling,'' \emph{arXiv preprint arXiv:2303.03926}, 2023.

\bibitem{desplanques20_interspeech}
B.~Desplanques, J.~Thienpondt, and K.~Demuynck, ``{ECAPA-TDNN: Emphasized Channel Attention, Propagation and Aggregation in TDNN Based Speaker Verification},'' in \emph{Proc. Interspeech 2020}, 2020, pp. 3830--3834.

\end{thebibliography}

\end{document}